\documentclass[twoside,twocolumn, 10pt]{article}

\usepackage{amsmath}
\usepackage{url}
\usepackage{subcaption}
\usepackage{array,multirow,graphicx}
\usepackage{authblk}
\usepackage{amsmath,amssymb}
\usepackage{booktabs}

\usepackage[margin=0.7in,columnsep=20pt]{geometry}
\DeclareMathAlphabet{\pazocal}{OMS}{zplm}{m}{n}
\DeclareMathOperator{\EX}{\mathbb{E}}
\newcommand\figref{Fig.~\ref}
\newcommand\tabref{Table~\ref}
\newcommand\eref{Equation~\ref}

\newcommand\blfootnote[1]{%
  \begingroup
  \renewcommand\thefootnote{}\footnote{#1}%
  \addtocounter{footnote}{-1}%
  \endgroup
}

\title{Video-Driven Speech Reconstruction using Generative Adversarial Networks}
\author[1,2]{Konstantinos Vougioukas}
\author[1]{Pingchuan Ma}
\author[1,2]{Stavros Petridis}
\author[1,2]{Maja Pantic}
\affil[1]{iBUG Group, Imperial College London}
\affil[2]{Samsung AI Centre, Cambridge, UK}

\begin{document}
\maketitle
\begin{abstract}
Speech is a means of communication which relies on both audio and visual information. The absence of one modality can often lead to confusion or misinterpretation of information. In this paper we present an end-to-end temporal model capable of directly synthesising audio from silent video, without needing to transform to-and-from intermediate features. Our proposed approach, based on GANs is capable of producing natural sounding, intelligible speech which is synchronised with the video. The performance of our model is evaluated on the GRID dataset for both speaker dependent and speaker independent scenarios. To the best of our knowledge this is the first method that maps video directly to raw audio and the first to produce intelligible speech when tested on previously unseen speakers. We evaluate the synthesised audio not only based on the sound quality but also on the accuracy of the spoken words.

\end{abstract}

\section{Introduction}
\label{sec:intro}
\blfootnote{* The first author performed this work during his internship at Samsung}
Lipreading is a technique that involves understanding speech in the absence of sound, primarily used by people who are deaf or hard-of-hearing. Even people with normal hearing depend on lip movement interpretation, especially in noisy environments.

The ineffectiveness of audio speech recognition (ASR) methods in the presence of noise has lead to the research of automatic visual speech recognition (VSR) methods. Thanks to recent advances in machine learning VSR systems are capable of performing this task with high accuracy \cite{assael2016lipnet, petridis2017end, petridis2018end, stafylakis2017combining,chung2016lipSentences}. 
Most of these systems output text but there are many applications such as videoconferencing in noisy or silent environments that would benefit from the use of video-to-speech systems.

\begin{figure}[t]
  \centering
  \includegraphics[width=0.78\linewidth]{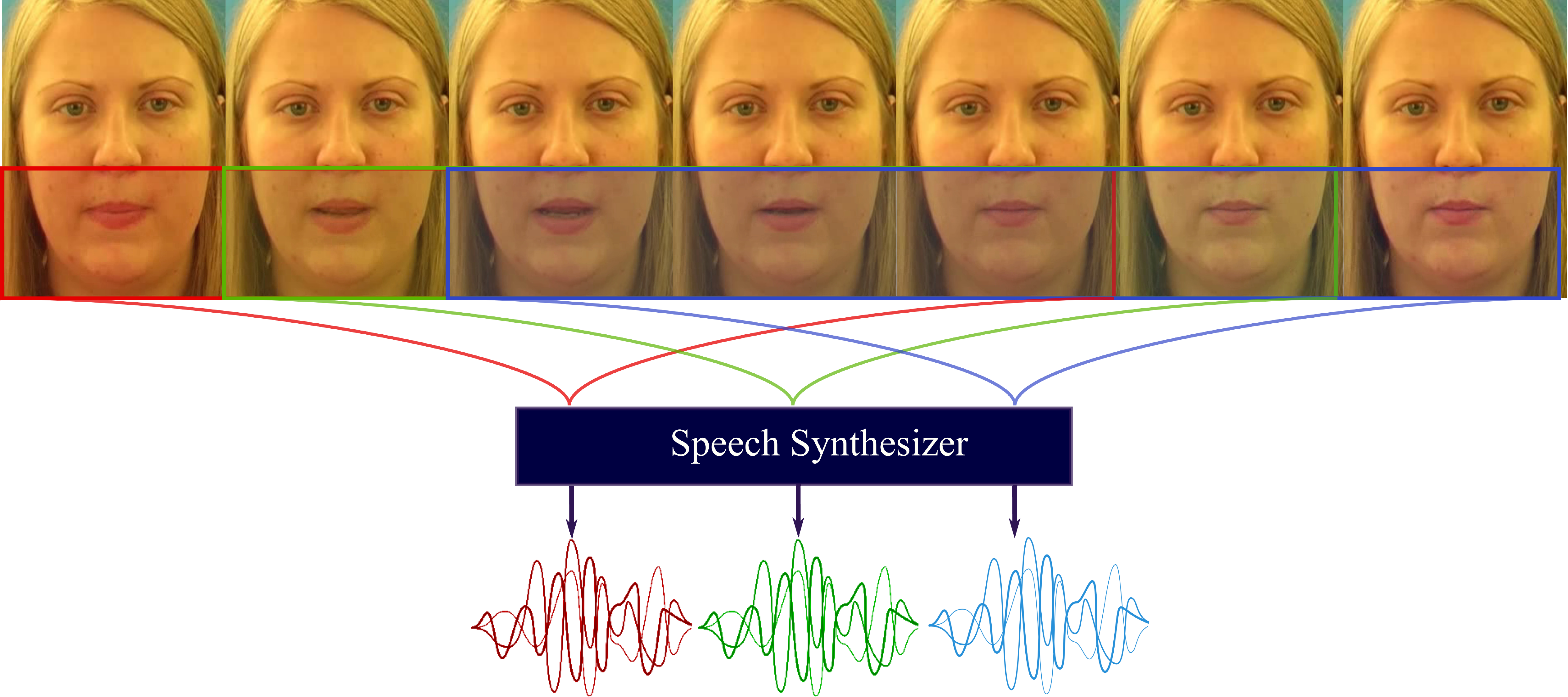}
  \caption{The speech synthesizer accepts a sequence of frames and produces the corresponding audio sequence.}
  \label{fig:speech_production}
\end{figure}

One possible approach for developing video to speech systems is to combine VSR with text-to-speech systems (TTS), with text serving as an intermediate representation. However, there are several limitations to using such systems. Firstly, text-based systems require transcribed datasets for training, which are hard to obtain because they require laborious manual annotation. Secondly, generation of the audio can only occur at the end of each word, which imposes a delay on the pipeline making it unsuitable for real-time applications such as video-conferencing. Finally, the use of text as an intermediate representation prevents such systems from capturing emotion and intonation, which results in unnatural speech. 

Direct video-to-speech methods do not have these drawbacks since audio samples can be generated with every video frame that is captured. Furthermore, training of such systems can be done in a self-supervised manner since most video comes paired with the corresponding audio. For these reasons video-to-speech systems have been recently considered.

Such video-to-speech systems are proposed by Le Cornu and Miller in  \cite{cornu2015reconstructing, cornu2017generating}, which use Gaussian Mixture Models and Deep Neural Networks (DNN) respectively to estimate audio features, which are fed into a vocoder to produce audio, from visual features. However, the hand-crafted visual features used in this approach are not capable of capturing the pitch and intonation of the speaker and in order to produce intelligble results they have be artificially generated. 


Convolutional neural networks (CNNs) have been shown to be powerful feature extractors for images and videos and have replaced handcrafted features in more recent works. One such system is proposed in \cite{ephrat2017vid2speech} to predict line spectrum pairs (LSPs) from video. The LSPs are converted into waveforms but since excitation is not predicted the resulting speech sounds unnatural. This method is extended in \cite{ephrat2017improved} by adding optical flow information as input to the network and by adding a post-processing step, where generated sound features are replaced by their closest match from the training set. A similar method that uses multi-view visual feeds has been proposed in \cite{kumar2018lipper}. Finally, Akbari et. al. \cite{akbari2018lip2audspec} propose a model that uses CNNs and recurrent neural networks (RNNs) to transform a video sequence into audio spectrograms, which are later transformed into waveforms using the algorithm proposed in \cite{chi2005multi}.

\begin{figure*}[t!]
  \centering
  \includegraphics[width=0.9\linewidth]{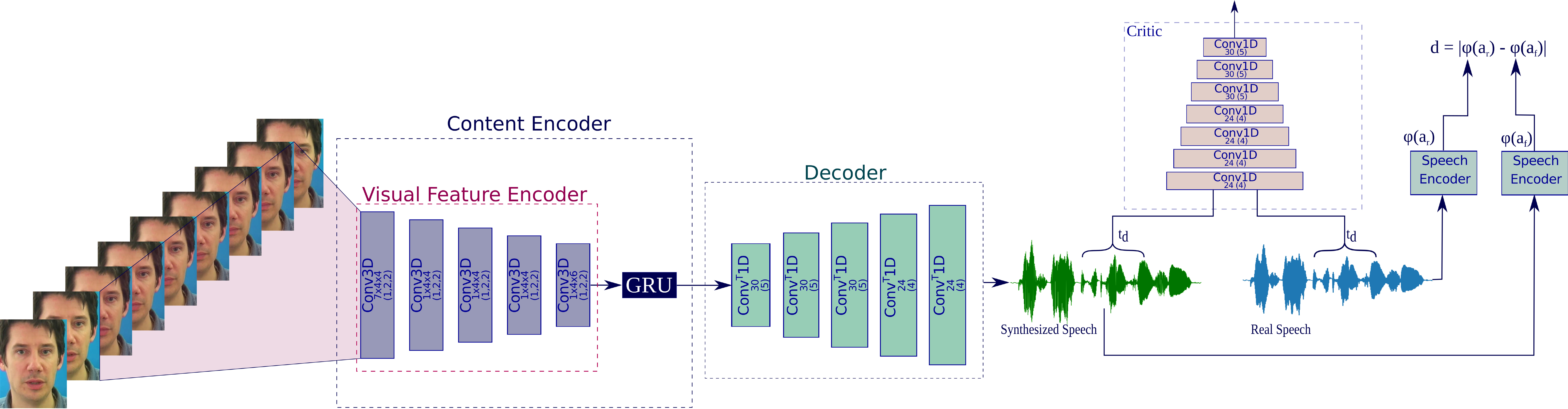}
  \caption{Architecture for the generator network consisting of a content encoder and an audio frame decoder. Convolutional layers are represented with Conv blocks with the kernel size depicted as $k_{time} \times k_{height} \times k_{width}$ and with strides for the respective axes appearing in parentheses. The critic accepts as input samples of $t_d=1s$ and determines whether they come from real or generated audio.}
  \label{fig:model_block}
\end{figure*}

In this work, we propose an end-to-end model that is capable of directly converting silent video to raw waveform, without the need for any intermediate handcrafted features as shown in \figref{fig:speech_production}. Our method is based on generative adversarial networks (GANs), which allows us to produce high fidelity (50KHz) audio sequences of realistic intelligible speech. Contrary to the aforementioned works our model is capable of generating intelligible speech even in the case of unseen speakers\footnote{Generated samples online at the following website: \\ \url{https://sites.google.com/view/speech-synthesis/home}}. The generated speech is evaluated using standard audio reconstruction and speech synthesis metrics.  Additionally, we propose using a speech recognition model to verify the accuracy of the spoken words and a synchronisation method to quantify the audio-visual synchrony.

\section{Video-driven Speech Reconstruction}
\label{sec:model}
The proposed model for speech reconstruction is made up of 3 sub-networks. The generator network, shown in \figref{fig:model_block} is responsible for transforming the sequence of video frames into a waveform. During the training phase the critic network drives the  generator to produce waveforms that sound similar to natural speech. Finally, a pretrained \textit{speech encoder} is used to conserve the speech content of the waveform. 

\subsection{Generator}
The generator is made up of a \textit{content encoder} and an \textit{audio frame decoder}. The \textit{content encoder} consists of a \textit{visual feature encoder} and an RNN. The \textit{visual feature encoder} is a 5 layer 3D CNN responsible for encoding information about the visual speech which is present in a window of $N$ consecutive frames. These encodings $z_s$, which are produced at each time step are fed to a single layer gated recurrent unit (GRU) network which produces a series features $z_c$ describing the content of the video. The \textit{audio frame decoder} receives these features as input and generates the corresponding window of audio samples. Batch normalization \cite{ioffe2015} and ReLU activation functions are used throughout the entire generator network except for the last layer in the \textit{visual feature encoder} and \textit{decoder}, where the hyperbolic tangent non-linearity is used without batch normalization. 

\subsection{Critic}
\label{sec:disc}
The critic is a 3D CNN which is given audio clips of fixed length $t_d$ from the real and generated samples. During training the critic learns a 1-Lipschitz function $D$ which is used to calculate the Wasserstein distance between the distribution of real and generated waveforms. In order to enforce the Lipschitz continuity on $D$ the gradients of  critic's output with respect to the inputs are penalized if they have a norm larger than 1 \cite{gulrajani2017improved}. The penalty is applied independently to all inputs and therefore batch normalization should not be used in any layers of the critic since it introduces correlation between samples of the batch. The audio clips which are provided as input to the critic are chosen at random from both the real and generated audio sequences using a uniform sampling function $S$.

\subsection{Speech Encoder}
The \textit{speech encoder} network is used to extract speech features from the real and generated audio. This network is taken from the pretrained model proposed in \cite{Vougioukas2018End-to-EndGANs}, which performs speech-driven facial animation. Similarly to our model this network is trained in a self-supervised manner and learns to produce encodings that capture speech content that can be used to animate a face. Using this module we are able to enforce the correct content onto the generated audio clips through a perceptual loss.

\subsection{Training}
Our model is based on the Wasserstein GAN proposed in \cite{gulrajani2017improved}, which minimises the Wasserstein distance between the real and fake distribution. Since optimising the Wasserstein distance directly is intractable the Kantorovich-Rubinstein duality is used to obtain a new objective \cite{villani2008optimal}. The adversarial loss of our model is shown in \eref{eq:adv_loss}, where $D$ is the critic function, $G$ is the generator function, $x$ is a sample from the distribution of real clips $P_r$ and $\widetilde{x}$ is a sample from the distribution of generated clips $P_g$. The gradient penalty shown in \eref{eq:gp} is calculated with respect to the input $\hat{x}$ sampled from distribution $P_{\hat{x}}$, which contains all linear interpolates between $P_r$ and $P_g$.

\begin{equation}
    \pazocal{L}_{adv} =  \EX_{x \sim P_r} [D(S(x))] - \EX_{\widetilde{x} \sim P_g}[D(S(\widetilde{x}))]
    \label{eq:adv_loss}
\end{equation}

\begin{equation}
    \pazocal{L}_{gp} =  \EX_{\hat{x} \sim P_{\hat{x}}} [(||\nabla_{\hat{x}}D(\hat{x})||_2 - 1)^2]
    \label{eq:gp}
\end{equation}

The \textit{speech encoder} maps a waveform $x$ to a feature space through a function $\phi$. During training a perceptual loss, corresponding to the $L_1$ distance between the features obtained from mapping real and generated audios is minimized. This forces the network to learn high-level features correlated with speech. The perceptual loss is shown in \eref{eq:perceptual}. 

\begin{equation}
    \pazocal{L}_p = \left|\phi(x)-\phi(\widetilde{x}) \right|
    \label{eq:perceptual}
\end{equation}

An $L_1$ reconstruction loss is also used to ensure that the generated waveform closely matches the original. Finally, we use a total variation (TV) regularisation factor in our loss described in \eref{eq:tv loss}, where $T$ is the number of samples in the audio. This type of regularization penalizes non-smooth waveforms and thus reduces the amount of high frequency noise in the synthesized samples.

\begin{equation}
    \pazocal{L}_{TV} = \frac{1}{T}\sum\limits_{t}\left|\widetilde{x}_{t+1}-\widetilde{x}_t \right|
    \label{eq:tv loss}
\end{equation}

The final objective used to obtain the optimal generator is a combination of the above losses as shown in \eref{eq:total loss}. The loss factors are weighted so that each loss has approximately equal contribution. The weights we used were $\lambda_{L_1}=150$, $\lambda_{tv}=120$, $\lambda_{gp}=10$ and $\lambda_{p}=70$.

\begin{equation}
    \pazocal{L} =\underset{G}{\min}\underset{D}{\max}\pazocal{L}_{adv} + \lambda_{L_1}\pazocal{L}_{L_1} + \lambda_{TV}\pazocal{L}_{TV} + \lambda_{p}\pazocal{L}_{p} + \lambda_{gp}\pazocal{L}_{gp}
    \label{eq:total loss}
\end{equation}

When training a Wasserstein GAN the critic should be trained to optimality \cite{arjovsky2017wasserstein}. We therefore perform 6 updates on the critic for every update of the generator. We use the Adam \cite{Kingma2014} optimizer with a learning rate of 0.0001 for both the generator and critic and train until no improvement is seen on the mel-cepstral distance between the real and generated audio from the validation set for 10 epochs. We use a window of $N=7$ frames as input for the generator and input a clip of $t_d=1s$ for the critic.


\section{Experimental Setup}
Our model is implemented in Pytorch and takes approximately 4 days to train on a Nvidia GeForce GTX 1080 Ti GPU. During inference our model is capable of generating audio for a 3s video recorded at 25 frames per second (fps) in 60ms when running on a GPU and 6s when running on the CPU.

\subsection{Dataset}
\label{sec:dataset}
The GRID dataset has 33 speakers each uttering 1000 short phrases, containing 6 words taken from a limited dictionary. The structure of a GRID sentence is described in \tabref{tab:gridstructure}. We evaluate our method in both speaker dependent and speaker independent settings. Subjects 1, 2, 4 and 29 were used for the subject dependent task and videos were split into training, validation and test sets using a random 90\%-5\%-5\% split respectively. This setup is similar to that used in \cite{akbari2018lip2audspec}. 

In the subject independent setting the data is divided into sets based on the subjects. We use 15 subjects for training, 8 for validation and 10 for testing. We use the split proposed in \cite{Vougioukas2018End-to-EndGANs}.

\begin{table}[h!]
\caption{Structure of a typical sentence from the GRID corpus.}
\centering
\footnotesize
\begin{tabular}{cccccc}
\toprule
 Command& Color& Preposition& Letter& Digit & Adverb	\\
\midrule
bin 	&  blue		& at  	&  A-Z	                & 0-9	&   again    \\
lay 	&  green 	& by  	&  \textbackslash \{W\} &   	& now      	 \\
place	&  red		& in 	&			            & 		& please 	 \\
set		&  white	& with 	& 		                & 		& soon		 \\
\bottomrule \\
\end{tabular}
\normalsize
\label{tab:gridstructure}
\end{table}

As part of our preprocessing all faces are aligned to the canonical face using 5 anchor points taken from the edge of the eyes and the tip of the nose. Video frames are normalised, resized to 128x96 and cropped keeping only the bottom half, which contains the mouth. Finally, data augmentation is performed by mirroring the training videos.

\subsection{Metrics}
\label{sec:metrics}
Evaluation of the generated audio is not a trivial task and there is no single metric capable of assessing all aspects of speech such as quality, intelligibility and spoken word accuracy. For this reason we employ multiple methods to evaluate the different aspects of our generated samples. In order to measure the quality of the produced samples we use the mean mel-cepstral distortion (MCD) \cite{kubichek1993mcd}, which measures the distance between two signals in the mel-frequency cepstrum and is commonly used to assess the performance of speech synthesizers. We also use Short Term Objective Intelligibility (STOI) metric which measures the intelligibility of the generated audio clip.

Speech quality is measured with the perceptual evaluation of speech quality (PESQ) metric \cite{rix2001pesq}. PESQ was originally designed to quantify degradation due to codecs and transmission channel errors. However, this metric is sensitive to changes in the speaker's voice, loudness and listening level \cite{rix2001pesq}. Therefore, although it may not be an ideal measure for this task we still use it in order to be consistent with previous works.


\begin{figure*}[t!]
\begin{subfigure}{.3\textwidth}
  \centering
  \includegraphics[width=.95\linewidth]{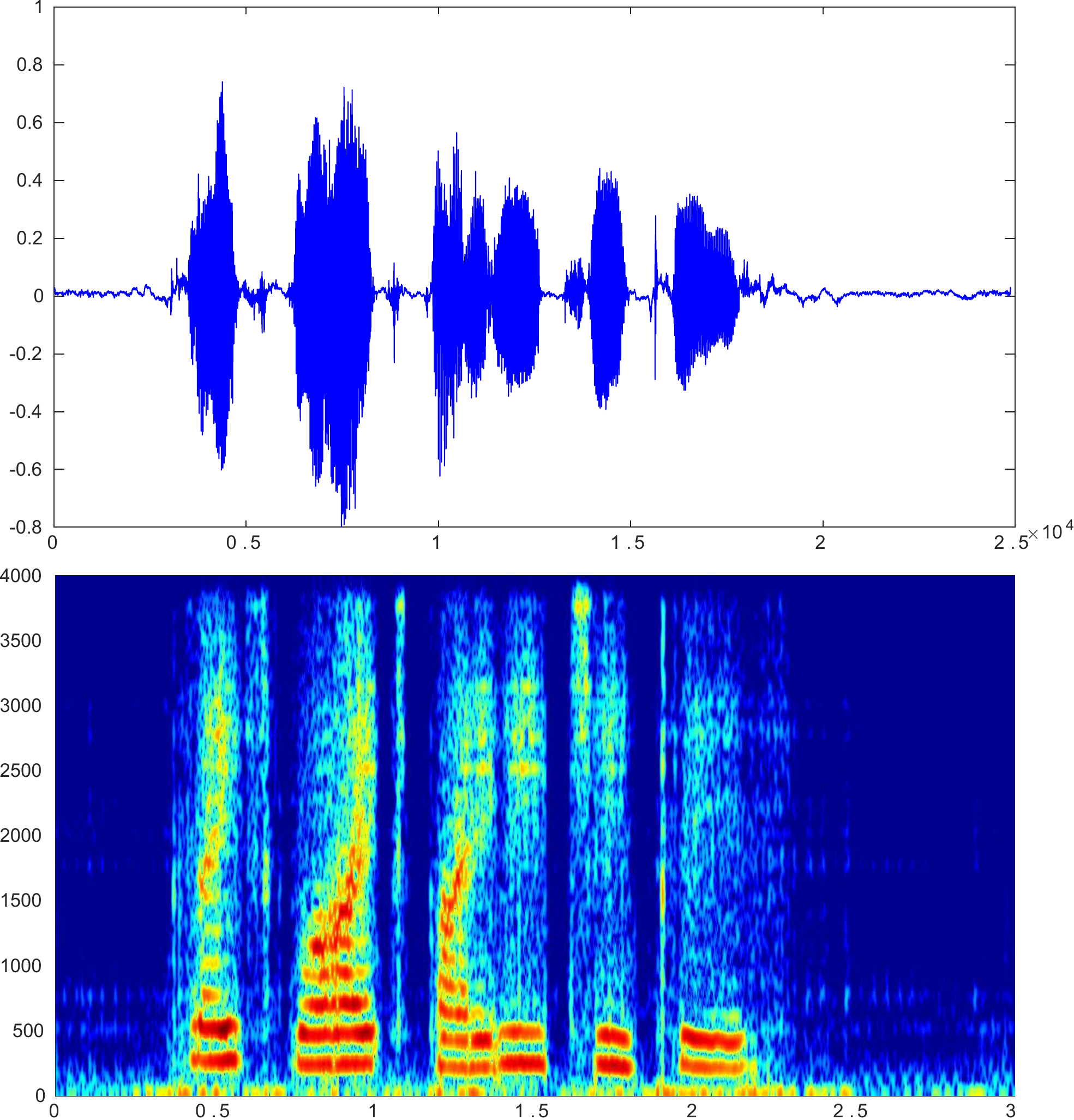}
  \caption{Proposed model}
  \label{fig:proposed_spectrogram}
\end{subfigure}
\begin{subfigure}{.33\textwidth}
  \centering
  \includegraphics[width=0.90\linewidth]{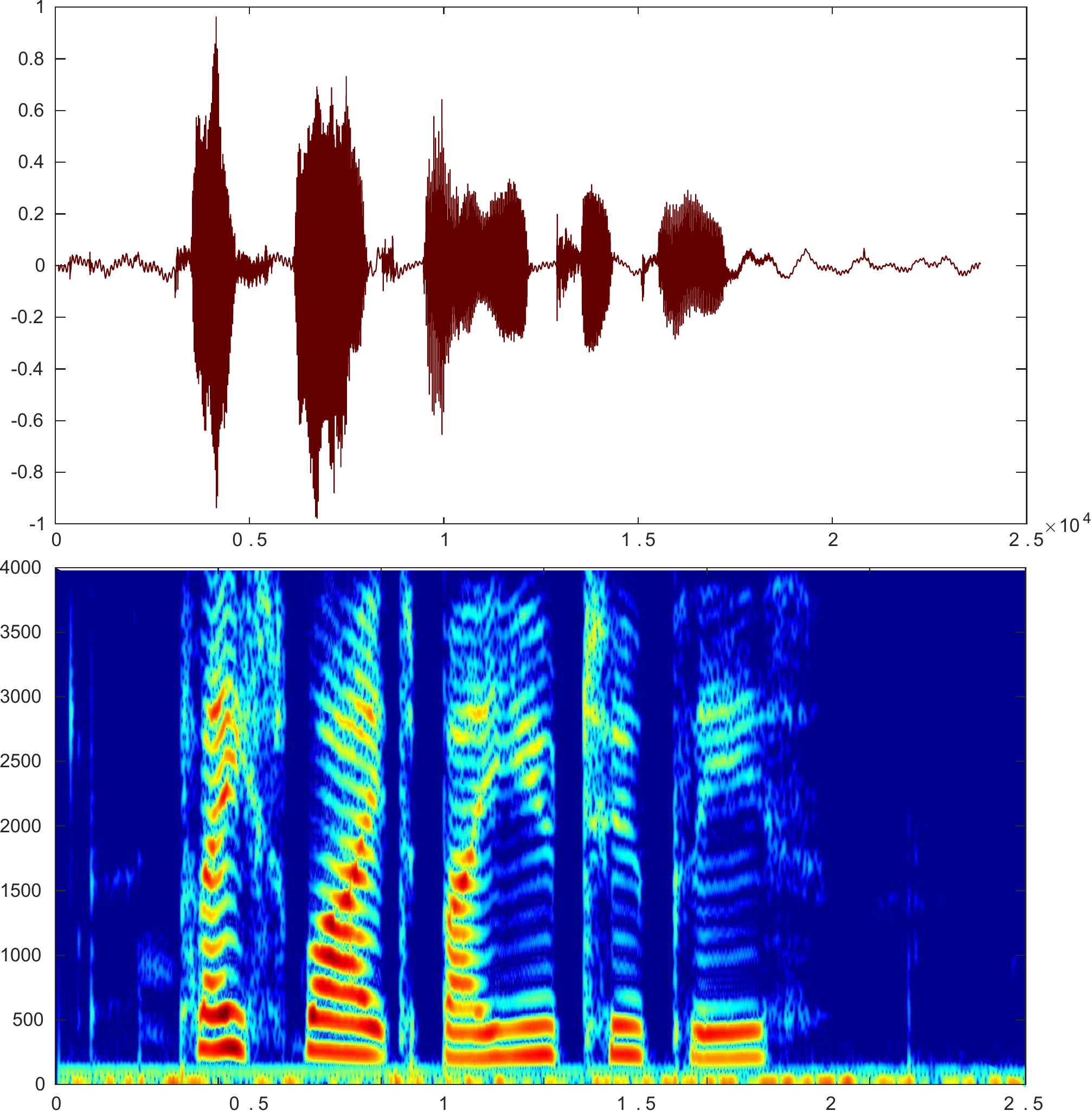}
  \caption{Real waveform}
  \label{fig:real_spectrogram}
\end{subfigure}
\begin{subfigure}{.33\textwidth}
  \centering
  \includegraphics[width=.91\linewidth]{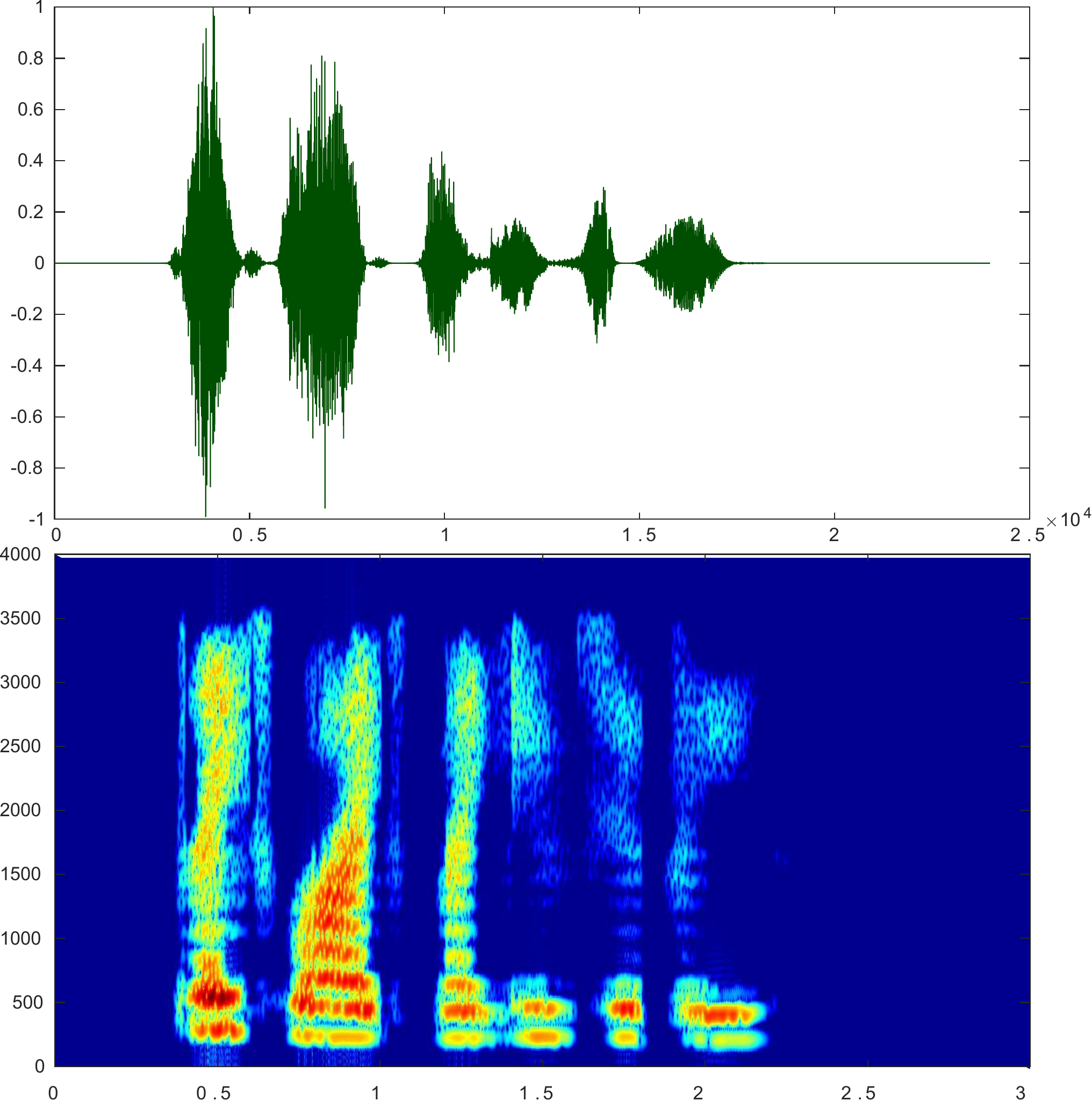}
  \caption{\textit{Lip2AudSpec}}
  \label{fig:lip2spec}
\end{subfigure}
\caption{Examples of real and generated waveforms and their corresponding spectrograms}
\label{fig:spectrograms}
\end{figure*}

\looseness - 1
In order to determine the synchronisation between the video and audio we use the pretrained SyncNet model proposed in \cite{Chung16a}. This model calculates the euclidean distance between the audio and video encodings for multiple locations in the sequence and produces the audio-visual offset (measured in frames) based on the location of the smallest distance and audio-visual correlation (confidence) based on the fluctuation of the distance.


Finally, the accuracy of the spoken message is also measured using the word error rate (WER) as measured by an ASR system, which is trained on the GRID training set. On ground truth audio the ASR system achieves 4\% WER.

\section{Results}
Our proposed model is capable ,of producing intelligible, high quality speech at high sampling rates such as 50KHz. We evaluate our model on the subject dependent scenario and compare it to the \textit{Lip2AudSpec} model proposed in \cite{akbari2018lip2audspec}. Furthermore we present results when the system is tested on unseen speakers. 

\subsection{Speaker Dependent Scenario}
\looseness -1
Examples of real and generated audio samples are compared in \figref{fig:spectrograms}. In order to present a fair comparison our audio is sub-sampled to match the rate of the sample produced by \textit{Lip2AudSpec}. Through inspection it is noticeable that our model captures more information than \textit{Lip2AudSpec} especially for very low and very high frequency content. This results in words being more clearly articulated when using our model. Our model introduces artifacts in the form of a low-power persistent hum in the waveform, which is also visible in the spectral domain.

We compare the samples produced by our model to those produced by \textit{Lip2AudSpec} using the metrics described in section \ref{sec:metrics}. The results are shown in \tabref{tab:metrics}. Our method out-performs \textit{Lip2AudSpec} in all the intelligibility tests (STOI, WER). Furthermore, our generated samples achieve better spectral accuracy as indicated by the smaller MCD. The \textit{Lip2AudSpec} achieves a better PESQ score which is likely due to the artifacts that are created using our model. Indeed, if we apply average filtering to the signal, which reduces these artifacts the PESQ increases to 1.80. However, this is done at the expense of sharpness and intelligibility since STOI drops to 4.7 and the WER increases to 36\%. Finally, both methods have similar scores in audio-visual synchrony, which is expected since they both use similar architectures to extract the visual-features.

\begin{table}[t]
  \caption{Metrics for the evaluation of the generated audio waveforms when testing on seen speakers. Best performance is shown in bold.}
  \label{tab:metrics}
  \centering
  \begin{tabular}{ c  c  c}
    \toprule
    Measure & \textit{Lip2AudSpec} & Proposed Model\\
    \midrule
    PESQ            & \textbf{1.82}          & 1.71\\
    WER             & 32.5\%            &  \textbf{26.6}\% \\
    AV Confidence   & 3.5               &  \textbf{4.4}\\
    AV Offset       & \textbf{1}             &  \textbf{1}  \\
    STOI            & 0.446             & \textbf{0.518} \\
    MCD             & 38.14             & \textbf{22.29} \\
    \bottomrule
  \end{tabular}
\end{table}

\begin{table}[t]
\footnotesize
  \caption{Ablation study performed in the subject dependent setting. Best performance is shown in bold.}
  \label{tab:ablation}
  \centering
  \begin{tabular}{cccccc}
    \toprule
    Model & PESQ & WER & AV Conf/Offset& STOI & MCD\\
    \midrule
    Full&\textbf{1.71}& \textbf{26.6\%} & \textbf{4.4(1)}& \textbf{0.518} & \textbf{22.29} \\
    w/o $\pazocal{L}_{L_1}$&1.45&33.2\%&3.9(1)&0.450&26.87\\
    w/o $\pazocal{L}_{TV}$&1.44 & 31.3 \% & 3.9(1)&0.483& 25,82\\
    w/o $\pazocal{L}_{p}$&1.14&83.3\%&2.0(5)&0.378&30.12\\
    \bottomrule
  \end{tabular}
  \normalsize
\end{table}

In order to quantify the effect of each loss term we perform an ablation study by removing one loss term at a time. The results of the ablation study are shown in \tabref{tab:ablation}. It is clear that removing each term makes the performance worse and the full models is the best over all performance measures. We note that the adversarial loss is necessary for the production of speech and when the system was evaluated without it generation resulted in noise. The other important contribution is made from the perceptual loss, without which the speech produced does not accurately capture the content of the speech. This is evident from the large increase in the WER, although this is reflected in the other metrics as well.

\subsection{Speaker Independent Scenario}
We present the results of our model on unseen speakers in \tabref{tab:metrics_si}. Comparison with other methods is not possible for this setup since other methods are speaker dependent (training the \textit{Lip2AudSpec} model on the same data did not produce intelligible results). From the results is evident that the speaker independent setting is a more complex problem. One of the reasons for this is the fact that the model can not learn the voice of unseen speakers. This results in generating a voice that does not correspond to the real speaker (i.e. female voice for a male speaker). Furthermore, in certain cases the voice will morph during the phrase. These factors likely account in large part for the drop in the PESQ metric, which is sensitive to such alterations. Morphing voices likely also affects the intelligibility of the audio clip, which is reflected in the WER, STOI and MCD. Finally we notice a slight improvement in audio visual synchrony which may be due to the increased number of samples seen during training.

It is important to note that the model has a different performance for each unseen subject. The WER fluctuates from 40\% to 60\% depending on the speaker. This is to be expected especially for subjects whose appearance differs greatly from the subjects in the training set. Additionally, since GRID is made up mostly of native speakers of English we notice that unseen subjects who are non-native speakers have worse WER.

\begin{table}[t]
\footnotesize
  \caption{Metrics for the evaluation of the generated audio waveforms when testing on unseen speakers}
  \label{tab:metrics_si}
  \centering
  \begin{tabular}{ c c c c c c}
    \toprule
    PESQ & WER   & AV Confidence & AV Offset & STOI     & MCD\\
    \midrule
    1.24 & 40.5 \% & 4.1 & 1  & 0.445 & 24.29\\
    \bottomrule
  \end{tabular}
  \normalsize
\end{table}

\section{Conclusions}
In this work we have presented an end-to-end model that reconstructs speech from silent video and evaluated in two different scenarios. Our model is capable of generating intelligible audio for both seen and unseen speakers. Future research should focus on producing more natural and coherent voices for unseen speakers as well as improving intelligibility. Furthermore, we believe that such systems are capable of reflecting the speakers emotion and should be tested on expressive datasets. Finally, a major limitation of this method is the fact that it operates solely on frontal faces. Therefore the natural progression of this work will be to reconstruct speech from videos taken in the wild.

\section{Acknowledgements}
We gratefully acknowledge the support of NVIDIA Corporation with the donation of the Titan V GPU used for this research and Amazon Web Services for providing the computational resources for our experiments.

\bibliographystyle{IEEEtran}
\bibliography{mybib}
\end{document}